 \documentclass[final,3p,times,twocolumn]{elsarticle}

\usepackage{graphicx}

\usepackage{amssymb}

\usepackage{siunitx}
\usepackage{amsmath}
\usepackage{amsthm}
\usepackage{mathtools}
\usepackage{xcolor}
\usepackage{commath}
\usepackage{amsfonts}
\usepackage{dsfont}
\usepackage{enumitem}
\usepackage{caption}
\usepackage{makecell}
\usepackage[printwatermark]{xwatermark}

\newcommand{\overbar}[1]{\mkern 1.5mu\overline{\mkern-1.5mu#1\mkern-1.5mu}\mkern 1.5mu}

\newcommand{\Z}{\mathbb{Z}}
\newcommand{\N}{\mathbb{N}}
\newcommand{\g}{\vec{g}}
\newcommand{\mean}[1]{{\left<#1\right>}}
\newcommand{\len}[1]{{l\left(#1\right)}}

\newcommand{\isneg}[1]{{\mathrm{neg}\left(#1\right)}}

\newcommand{\comment}[1]{}

\makeatletter
\newcommand*\bigcdot{\mathpalette\bigcdot@{.5}}
\newcommand*\bigcdot@[2]{\mathbin{\vcenter{\hbox{\scalebox{#2}{$\m@th#1\bullet$}}}}}
\makeatother
\newcommand\myeq{\mkern1.5mu{=}\mkern1.5mu}
\usepackage{physics}

\newcounter{bla}

\DeclareFontFamily{OMS}{oasy}{\skewchar\font48 }
\DeclareFontShape{OMS}{oasy}{m}{n}{%
         <-5.5> oasy5     <5.5-6.5> oasy6
      <6.5-7.5> oasy7     <7.5-8.5> oasy8
      <8.5-9.5> oasy9     <9.5->  oasy10
      }{}
\DeclareFontShape{OMS}{oasy}{b}{n}{%
       <-6> oabsy5
      <6-8> oabsy7
      <8->  oabsy10
      }{}
\DeclareSymbolFont{oasy}{OMS}{oasy}{m}{n}
\SetSymbolFont{oasy}{bold}{OMS}{oasy}{b}{n}

\DeclareMathSymbol{\smallleftarrow}     {\mathrel}{oasy}{"20}
\DeclareMathSymbol{\smallrightarrow}    {\mathrel}{oasy}{"21}
\DeclareMathSymbol{\smallleftrightarrow}{\mathrel}{oasy}{"24}
\newcommand{\tensor}[1]{\overset{\scriptscriptstyle\smallleftrightarrow}{#1}}

\newcommand{\Wc}{w}

\usepackage{hyperref}
\hypersetup{
	colorlinks=true,
	pdfstartview=FitV,
	linkcolor=blue,
	citecolor=blue,
	urlcolor=blue,
	pdfauthor={SCHUSTER Christian, christian.schuster@ipp.mpg.de},
	pdfsubject={Moment-Preserving and Mesh-Adaptive Reweighting Method for Rare-Event Sampling in Monte-Carlo Algorithms},
	pdftitle={Moment-Preserving and Mesh-Adaptive Reweighting Method for Rare-Event Sampling in Monte-Carlo Algorithms},
	plainpages=false,
}

\journal{Computer Physics Communications}

\begin{document}

\begin{frontmatter}



\title{Moment-Preserving and Mesh-Adaptive Reweighting Method for Rare-Event Sampling in Monte-Carlo Algorithms}


\author[a,b,c]{C. U. Schuster\corref{author}}
\author[b]{T. Johnson}
\author[a]{G. Papp}
\author[a]{R. Bilato}
\author[d]{S. Sipil\"a}
\author[d]{J. Varje}
\author[e]{M. Hasen\"ohrl}

\cortext[author] {Corresponding author.\\\textit{E-mail address:} christian.schuster@ipp.mpg.de}
\address[a]{Max-Planck-Institute for Plasma Physics, Garching, Germany}
\address[b]{Fusion Plasma Physics, EECS, KTH, Stockholm, Sweden}
\address[c]{Department of Physics, Technical University of Munich, Garching, Germany}
\address[d]{Department of Applied Physics, Aalto University, Aalto, Finland}
\address[e]{Department of Mathematics, Technical University of Munich, Garching, Germany}

\begin{abstract}
We present novel roulette schemes for rare-event sampling that are both structure-preserving and unbiased. The boundaries where Monte Carlo markers are split and deleted are placed automatically and adapted during runtime. Extending existing codes with the new schemes is possible without severe changes because the equation of motion for the markers is not altered. These schemes can also be applied in the presence of nonlinear and nonlocal coupling between markers. As an illustrative application, we have implemented this method in the ASCOT-RFOF code, used to simulate fast-ion generation by radio-frequency waves in fusion plasmas. In this application, with this method the Monte-Carlo noise level for typical fast-ion energies can be reduced at least of one order of magnitude without increasing the computational effort.

\end{abstract}


\end{frontmatter}
\section{Introduction}
Monte Carlo techniques can be efficient for solving a variety of problems \cite{hansmann1999new, dum1992monte, lemaire2005monte, saxton1994anomalous, brandimarte2014handbook, bal2011hybrid}. One example is plasma physics and the heating of plasmas with ion cyclotron resonance heating (ICRH) \cite{stixBook}. The particle distribution function $f$ in a plasma is governed by a Boltzmann equation in Fokker-Planck approximation
\begin{align}
\label{eq:fokkerPlanck}
	\dpd{f}{t} = - \vec{a} \, \nabla_{\vec{z}} f + \nabla_{\vec{z}} \cdot \left(\nabla_{\vec{z}} \cdot \frac{\tensor{b} \cdot \tensor{b}}{2} f \right).
\end{align}
$\vec{z}$ is the phase space coordinate, $\tensor{D} = \left(\tensor{b} \cdot \tensor{b}\right)/2$ is the diffusion tensor, and $\vec{a}$ the advection velocity. The single arrow denotes a vector, the double arrow a tensor. $\tensor{D}$ and $\vec{a}$ together are refered to as Fokker-Planck Coefficients (FPC). $f$ can be computed by simulating the corresponding Langevin equation \cite[p.~294ff]{ichimaru2018basic}
\begin{align}
\label{eq:langevin}
	\dif \vec{z} = \vec{a} \dif t + \tensor{b} \cdot \vec{\xi}(t) \sqrt{\dif t}
\end{align}
for test particles, which we call markers. $\vec{\xi}$ are Gaussian random numbers with zero mean and variance 1. The markers are distributed according to $f$, which can be reconstructed for example by using a histogram:
\begin{align}
\label{eq:histDef}
	\overbar{f}_\text{histogram bin}(t) = \frac{\int_\mathrm{bin} f(\vec{z},t) \dif \vec{z}}{\int_\mathrm{bin} \dif \vec{z}} \approx \frac{\sum_i \mathds{1}_\mathrm{bin}\left(\vec{z}_i(t)\right) W_i(t)}{\int_\mathrm{bin} \dif \vec{z}},
\end{align}
where we approximated the integral over the distribution function with the sum over the markers inside a bin of the histogram. $\mathds{1}_\mathrm{bin}(\vec{z})$ is 1 if $\vec{z}$ lies in the bin of the histogram, and 0 otherwise. $W_i$ is the weight of the $i$-th marker, which is constant unless reweighting is employed\footnote{Some schemes introduce $f(\vec{z})=g(\vec{z})\cdot w(\vec{z})$ \cite{brunner1999collisional}, and evolve the markers according to the Langevin equation for $g$. Each marker then has a weight which evolves in time and effectively gives $w(\vec{z})$. We instead directly solve for $f$. The weights are only adapted in discrete steps when we employ reweighting, which will be introduced later. The weights do not follow an equation of motion.}. The markers are randomly initialized such that equation \eqref{eq:histDef} is valid for the initial condition for $f$.

For ICRH the FPC have two contributions: collisions, for which a Maxwellian background plasma with which the particles collide can be assumed \cite{hirvijoki2014ascot}, and radio frequency (RF) interactions. 
The RF FPC depend on the amplitude of the wave field $A_\mathrm{wave}$. The wave amplitude is set by the injected power:
\begin{align}
	\label{eq:waveAmplitude}
	P_\mathrm{rf} = A_\mathrm{wave} \int _D g\left(f(z), \tensor{D}(z), v(z)\right) \, \mathrm{d}z,
\end{align}
where $A_\mathrm{wave} \cdot g$ is the density of the absorbed power, and $D$ the domain of the Fokker-Planck equation. When modeling an experiment one has to adjust the FPC such that the absorbed power is close to the power injected in the experiment. Equations \eqref{eq:fokkerPlanck} and \eqref{eq:langevin} are then nonlocal, complicating the numerical solution of equation \eqref{eq:fokkerPlanck}. 
The Langevin equation \eqref{eq:langevin} can still be solved for fusion plasmas by orbit following codes such as ASCOT \cite{hirvijoki2014ascot}. The code library RFOF \cite{johnson2011library} is used here to include the RF interaction.

A direct numerical solution of equation \eqref{eq:langevin} quickly becomes computationally expensive when the solution of the diffusion-advection equation varies by several orders of magnitude in the domain of the solution. As $f$ is proportional to the probability distribution of marker positions, only few markers are in areas where $f$ is small. Those areas are therefore only poorly resolved. Plasmas heated by ICRH exhibit a population of highly energetic ions, which are few compared to thermal ions but of prime interest \cite{kazakov2017efficient, weiland2017phase, killeen2012computational, heidbrink2008basic}. 

From importance sampling we know how to reduce the error of the simulation results in such regions of interest: We have to place many markers there. Equation \eqref{eq:histDef} tells us that this corresponds to lower marker weights in those regions compared to regions of less interest. As the simulation progresses, the markers will mix due to the stochastic term in equation \eqref{eq:langevin}, i.e. small markers will leave the region of interest and large markers will enter it. We need to adjust the weights of markers entering and leaving the region to keep the weight of markers in the region of interest smaller than the average weight, which is called reweighting. There exist different reweighting techniques in the literature which however suffer from drawbacks. The classic splitting and Russian Roulette techniques \cite[p. 99f]{hammersley2013MCmethods} lead to growing fluctuations of the marker number and total weight because the Russian Roulette does not conserve the conserved quantities of the underlying Fokker-Planck equation. For long times all markers will eventually be deleted (cf. the Bienaym\'e-Galton-Watson process \cite[5.4]{GeoffreyGrimmett2001}, which we introduce in Section \ref{russianRoulette}). There are other methods, based on merging markers \cite{martin2016octree}, which however bias the result. The mean result is then no longer the exact result of the underlying Fokker-Planck equation, although the extent of this effect can be mitigated by selecting well the markers that are merged (Octree binning) \cite{martin2016octree}.

There exist other methods for resolving regions of phase space with low densities, often collectively reffered to as rare event sampling. Here we however choose splitting and Russian Roulette techniques because
\begin{itemize}
	\item They can cope with the nonlinear and nonlocal property introduced by the fixed RF power (equation \eqref{eq:waveAmplitude}).
	\item One can avoid additional bias compared to the direct solution of the Langevin equation \eqref{eq:langevin} with the new splitting and roulette techniques presented in this paper. We show this analytically for a special case in in \ref{unalteredMean}, and verify it numerically \cite{schuster2019MscThesis}.
	\item The new reweighting schemes are structure preserving because no errors in the moments of the distribution function are introduced that grow as the simulation progresses.
	\item It is possible to extend existing codes that directly simulate the Langevin equation \eqref{eq:langevin} without big changes because the equation of motion for the markers is not altered.
\end{itemize}

This paper is structured as follows: In Sections \ref{splitting} and \ref{russianRoulette} we introduce the traditional splitting and roulette methods as tools for variance reduction. In Sections \ref{corrRoulette} - \ref{nonintegerWght} we present our new reweighting schemes, whose parameters can be chosen as described in Section \ref{parameterChoice}. The schemes are applied in Section \ref{results}. In Section \ref{ascotSection} we show how the new schemes can reduce by orders of magnitude the number of markers needed to obtain a certain noise amplitude for simulations with the orbit-following code ASCOT \cite{hirvijoki2014ascot}.

\section{Reweighting Techniques}
\label{reweighting}
Diffusion-advection equations are often solved by markers evolving according to the associated Langevin equation \eqref{eq:langevin}. This can be inefficient when the distribution function $f$ varies by orders of magnitude in its domain. When we calculate $f$ from the marker positions by using a histogram, we can use bin statistics\footnote{For bin statistics we assume that the markers are independent, and that the probability for a marker to be inside a histogram bin is small compared to the probability of being outside of the bin. The number of markers in the bin is then Poisson distributed, therefore the mean equals the variance.} to estimate the accuracy of the result.

When we assume that the number of markers in a bin is Poisson distributed, we obtain for the relative statistical error of the reconstructed $f$
\begin{align}
\label{eq:eRelNoRewei}
E_\mathrm{rel} = \frac{\sigma(\overbar{f})}{\left<\overbar{f}\right>} &= \frac{1}{\sqrt{\left<n\right>}},
\end{align}
where $\overbar{f}$ is the reconstructed $f$ from equation \eqref{eq:histDef}, $\sigma$ the standard deviation and $\left<n\right>$ is the mean number of markers in the corresponding bin of the histogram. When $f$ is small in a region of phase space compared to elsewhere in phase space, $\left<n\right>$ will also be small by definition. $E_\mathrm{rel}$ is then large. 

We can reduce $E_\mathrm{rel}$ by using more markers in the low density region. When we use more markers we have to reduce their weight accordingly to obtain the correct $f$. For a problem such as ICRF heating using differently weighted markers in different regions of phase space is not trivial because the simulated time scales are much larger than the collisional time scales. The markers will mix and the advantage of differently weighted markers will vanish. It is therefore necessary to reweight the markers, i.e. adjust their weights during runtime, as they move in the domain. This can be done using splitting techniques \cite[p. 100]{hammersley2013MCmethods} for markers that enter a region of lower weight, and employing Russian Roulette techniques for markers leaving low weight regions \cite[p. 99]{hammersley2013MCmethods}. For this we need to define discrete regions in phase space.

\subsection{Adaptive Weight Regions}
To decide how to reweight the markers we define nested (hyper)volumes in phase space, called weight regions or simply regions, each with a different weight for the markers therein. We define the weight regions with the help of a scalar function $\Wc(z)$. The weight region $i$ with weight $W_i$ is defined as $\left\{z \in D: W_i \geq \Wc(z) > W_{i+1}\right\}$, where $D$ is the whole phase space. The weight region for each marker $j$ and time step is found by computing $\Wc(z_j)$. Note that we can use a single index to identify weight regions in a multidimensional phase space by virtue of the mapping from phase space to $\mathbb{R}$ provided by $\Wc(z)$. The weights of the regions $W_i$ are not determined automatically and are free parameters.

The ratio between the weights of two adjacent regions $i$ and $i+1$
\begin{align}
R_{i,i+1} = \frac{W_{i}}{W_{i+1}}
\end{align}
is, for now, required to be an integer. After discussing the novel schemes in Sections \ref{corrRoulette} and \ref{detRoulette} we will have the tools to relax this constraint. 

By setting up weight regions we obtain the following important properties:
\begin{enumerate}[label=\arabic*)]
\item Markers at the same phase space position have the same weight. This is optimal, assuming uncorrelated markers, in the sense that the variance of equation \eqref{eq:histDef} is minimized. With reweighting schemes it can be beneficial to allow some mixing, as discussed in Section \ref{hyst}.
\item The largest weight that markers can obtain is given by the weight of the first weight region. There is, by design, no population of markers with higher weights than desired.
\item We can freely choose which regions of phase space to resolve with which weights. As we discuss in the next paragraphs we can even automatically adapt the regions during runtime.
\item Determining the weight region a marker belongs to is efficient. One only has to evaluate $\Wc$ and determine the largest $i$ with $W_i \geq \Wc$.
\end{enumerate}

As described above the weight of a marker is a function of its phase space position, and it would therefore not be necessary to store the weight. However, in Section \ref{hyst} we allow markers of different weight to mix to some extent, making it necessary to store the weight.

\label{weightRegions}
We should not define the weight regions arbitrarily. Bin statistics and importance sampling suggest that we should concentrate all markers in the region of phase space we are interested in. This consideration is, however, unaware of the fact that we still have to resolve the rest of phase space with sufficient resolution to obtain the boundary conditions for our regions of interest. We will later refer to the influx of markers from a poorly resolved region to a well resolved region as noisy flux.

One approach to sufficiently resolve the whole phase space is to keep the marker density in phase space approximately constant. In this case $\Wc  \propto f$, and we can determine $\Wc $ by calculating the binned density whenever we reweight. $\Wc $ is not constant in time. This rather radical approach might reduce the accuracy of regions with large $f$ considerably compared to simulations with equally weighted markers. As presented in detail earlier \cite{schuster2019MscThesis} we can interpolate between $\Wc(z, t) = \mathrm{const}$ and $\Wc \propto f(z,t)$. The idea is that the accuracy depends on the phase space density of markers, and by interpolating linearly between the marker densities we can interpolate freely between the weighting functions. In Section \ref{flow} and Figure \ref{fig:interpWghtFctB04} we will show that such an interpolated weighting function can be used to resolve both low and high density regions well. We can write the interpolated weight function as

\begin{align}
\label{eq:combinedWghtFct}
\Wc (z,t) = \left[M \left( \frac{1-a}{N} + \frac{a}{V_D f(z,t)} \right)\right] ^{-1},
\end{align}
where 
\begin{align*}
N &\coloneqq \int _D f(z) \, \mathrm{d}z \\
V_D &\coloneqq \int _D \mathrm{d}z
\end{align*}
are constants, and $M$ and $a$ are parameters that can be freely chosen. $M$ is the (mean) number of markers used in the simulation, and $a \in [0,1]$ determines how the weight regions are placed. $a=0$ yields markers of equal weight, and $a=1$ a spatially constant marker density in phase space. As long as $N$ and $V_D$ are conserved, the mean number of markers does not change during the simulation, provided the discretized weights $W_i$ are close to one another and can be considered continuous. If $V_D$ is infinite we can for example define a minimum weight beyond which markers are no longer split. Then the mean number of markers can change during the simulation even when $N$ is conserved.

Calculating $\Wc $ according to equation \eqref{eq:combinedWghtFct} in general requires communication between threads because we need to evaluate the local $f$. Usually the distribution function evolves on longer time scales than the time between reweightings. Then $\Wc $ does not need to be calculated every time, allowing for reweighting without communication between threads for each operation. Another implementation detail is the choice of $W_i$, which is explained in section \ref{parameterChoice}.

\subsection{Splitting}
\label{splitting}
When we split one marker into $R$ markers we have to
\begin{enumerate}[label=\arabic*)]
\item reduce the weight of the marker by a factor $1/R$
\item create $R-1$ identical copies of the marker.
\end{enumerate}

Immediately after splitting there is no improvement of the accuracy. The $R$ markers at the same position are mathematically identical to the original marker, but reqire $R$-times as many computations to advance. The accuracy however improves as the simulation progresses: The stochastic motion due to the diffusive term in equations \eqref{eq:fokkerPlanck} and \eqref{eq:langevin} is different for each marker. After long times the accumulated stochastic motion is sufficiently large that the marker position is mostly independent of the initial position. We call the time scale required for this to happen the decorrelation time. After the markers are decorrelated we can treat them as independent, and equation \eqref{eq:eRelNoRewei} holds again. 

Observe that the notion of independent markers depends on what we evaluate with the markers. When considering the binned density, as used for histograms, the markers can be considered independent when the $R$ intially identical markers are not all in the same bin. This happens when the markers have independently traveled one bin-length, which is therefore the length scale on which the markers become independent. To summarize, markers decorrelate on shorter lengths when the measure we evaluate, e.g. the binned density, is more localized. The decorrelation is apparent in figure \ref{fig:wiener2d}, where the variance of the result drops to the prediction assuming uncorrelated markers approximately at a distance of 0.2 away from the boundary where markers are split.

\subsection{Russian Roulette}
\label{russianRoulette}
When markers travel from the low-weight region $i+1$ to the high-weight region $i$, we have to reduce their number and therefore increase their weight. Contrary to approaches in current literature \cite{vranic2015particleMergingAlgorithmPIC, martin2016octree} we do not merge markers as this will result in a biased estimate of the distribution function \cite{vranic2015particleMergingAlgorithmPIC}. Instead, we delete some of the markers. With the traditional so-called Russian Roulette \cite[p. 99]{hammersley2013MCmethods} we delete markers that travel from region $i+1$ to region $i$ with a probability of $1-1/R_{i,i+1}$, and multiply the weight of the survivors by $R_{i,i+1}$. If a marker jumps from region $i$ to $i+j$ or $i-j$ it is split $j$ times or undergoes the roulette $j$ times, depending on the sign. Note that even when a method is unbiased it can lead to fluctuations that reduce the accuracy of the simulation.

When implementing reweighting as so far described, i.e. we
\begin{itemize}
	\item define weight regions,
	\item split markers when they travel to a region of lower weight,
	\item delete markers with the Russian Roulette when the travel to a region of higher weight,
\end{itemize}
the number of markers will fluctuate and eventually ($\mathrm{lim} \,_{t\rightarrow \infty}$) all markers will be deleted: A marker of weight $W$ travels to a lower-weight region and is split into $R$ markers. Subsequetely these markers eventually return to the original region, meaning every marker is deleted with a probability of $1-1/R$. The number of markers $n$ is then a stochastic variable, with a variance $\sigma_0^2 \approx n$. As the simulation progresses there will be many subsequent splittings and roulettes. Neglecting correlations this is the Bienaym\'e-Galton-Watson process \cite[5.4]{GeoffreyGrimmett2001}. This insight yields two important conclusions:
\begin{itemize}
    \item The probability that all markers are deleted for $\mathrm{lim} \,_{t\rightarrow \infty}$ is 1.
    \item The variance of the marker number is $\sigma^2 = m \sigma_0^2$ after reweighting each marker $m$ times.
\end{itemize}
Let us consider mass conservation as an example: If we reweight each marker once for every marker in our simulation, we can expect that the error of the total mass is comparable to the total mass itself. For longer simulations we therefore have to take precautions against this effect. We now present two new reweighting schemes that are structure-conserving in the sense that the mass error does not grow as the simulation progresses.

\subsubsection{Correlated Roulette}
\label{corrRoulette}
From now on we will describe the new techniques to avoid the shortcomings of the traditional Russian Roulette. Our first approach is to use correlated instead of uncorrelated random numbers for the Russian Roulette. We require $R = k/l$ to be a rational number with $k$ and $l$ integers and $k\geq l$. To let, on average, only 1 out of $R$ markers survive the crossing to the high weight region we delete exactly $k-l$ out of every $k$ markers that travel to the high weight region. 

One way to implement this is to take a Boolean array with $k$ entries. The array contains $l$ entries of 1 and $k-l$ entries of 0, where 0 represents deleting a marker and 1 letting it survive. The order is randomly assigned. For the first marker that crosses the boundary we use the first entry to decide whether to delete it or not, for the second marker the second entry and so forth. After the last entry of the array has been used we fill the array with entries in a new and random order and start with the first entry for the next marker. Compared to an uncorrelated random number generator that gives 1 with probability $(k-l)/k$ and 0 otherwise we built a random number generator with the same probabilities for the results, but with correlations between the generated random numbers such that we always delete $(k-l)$ markers out of $k$. We use such a correlated random number generator for each boundary between weight regions.

After $k$ markers have traveled to the high weight region, the total weight is identical to the total weight before the markers traveled to the high weight region. The total weight can therefore only deviate slightly from the true value, and this deviation does not grow larger with longer simulated time. Also, this deviation of total weight is $\mathcal{O}(N^{-1})$\footnote{Suppose $n_0 \propto N$ markers travel from the low-weight to the high-weight region. They initially have weight $W_{i+1}$, after passing the roulette they have weight $W_i$. The number of markers that survive the roulette is $n_1$. The correlated roulette would ideally preserve the total weight 
\begin{align*}
w_\mathrm{tot,ideal} = n_0 \, W_{i+1} = n_{1,\mathrm{ideal}} \, W_i.
\end{align*}
An upper bound for the error of the roulette, i.e. the mismatch between $n_{1,\mathrm{ideal}}$ and $n_{1,\mathrm{real}}$, is k. Because $W_i \propto 1/N$ we obtain
\begin{align*}
\left|w_\mathrm{tot,ideal} - w_\mathrm{tot,real}\right| = \left| n_{1,\mathrm{ideal}} - n_{1,\mathrm{real}} \right| \, W_i \leq k \, W_i \propto 1/N. \; \blacksquare
\end{align*}}, with $N$ the number of markers used in the simulation, and quickly becomes negligible when $N\gg k$.

\subsubsection{Deterministic Roulette}
\label{detRoulette}
Our second approach aims not only at avoiding growing fluctuations of the zeroth moment of the distribution function but of all moments. While we will not prove this property rigorously, we will give an argument in the following paragraphs. In section \ref{wiener}, where we consider the binned distribution function, this advantage will become visible.

As an additional property of each marker we store the region in which the marker was created and use this information instead of random numbers for deciding which markers to delete. Markers are deleted when they travel to a region of higher weight than what they were created with. Markers that originated in the highest weight region will never be deleted. But as all other markers their weights change as the markers move between regions and generate new markers by splitting. All markers that are created by splitting eventually travel to a weight region where they are deleted. The marker population is therefore composed of two parts: 
\begin{enumerate}[label=\Alph*)]
\item Markers created in the highest weight regions. Their distribution is uninfluenced by reweighting, and therfore reweighting cannot introduce any fluctuations for them.
\item Markers created by the A-markers when they travel to a lower weight region. After an A-marker created a B-marker it needs to return to the larger-weight region before it can create another sucher B-marker. Because both A and B markers follow the same equation of motion, the probability for the A-marker to return to the high weight region until some time $\tau$ is the same as for the B-marker to be deleted before the time $\tau$. We therefore expect that there is no accumulation or depletion of B-markers as the simulation progresses because the whole population of B-markers is continuously renewed by the A-markers whose population is uninfluenced by reweighting. 
\end{enumerate}
We still introduce fluctuations by reweighting because markers that were created in the low weight regions and the original marker from the high weight region will in general not return simultaneously. But these fluctuations do not accumulate or grow as the simulation progresses. 

When we initialize the markers in the beginning of the simulation we have to set their creation region. A simple approach would be to create markers with weights for the highest weight region and reweight them. A method that however introduces fewer fluctuations and can generate a list of markers with properly set creation weights from any list of (non-uniformly weighted) markers is presented in \cite{schuster2019MscThesis}. This initialization method can also be used to add markers during runtime.

\subsection{Hysteresis Region}
\label{hyst}
Reweighting introduces fluctuations, even with the correlated and the deterministic roulette.

A Monte Carlo marker typically moves on a zigzag trajectory as known from Brownian motion. When a marker travels from one weight region to another we can therefore expect it to cross the boundary many times. If the time resolution becomes infinite we expect an infinite number of crossings because the underlying Wiener process is self-similar \cite{embrechts2009selfsimilar}.

Every time we reweight a marker we introduce fluctuations. It is therefore advisable to not reweight too often, e.g. after every time step. While it is not obvious what the optimal time between reweightings is, letting the markers mix until they have typically moved across a weight region before reweighting has proven to yield satisfying results \cite{schuster2019MscThesis}.

Selecting the corresponding time between reweightings can be problematic when the diffusion coefficient varies in phase space or when it is not explicitly known. In this case we can reweight after short times and define a hysteresis region where the markers are allowed to mix. Without hysteresis, we split a marker at position $z$ with weight $W_i$ when $\Wc (z) \leq W_{i+1}$, and subject it to the roulette when $\Wc (z) > W_{i}$. With a hysteresis of width $h$ we increase the region the marker can move in before being reweighted by exponentiating with $h/2$: $\left\{z \in D: W_i \, \left(r_{i-1,i}\right) ^ {h_{i-1,i}/2} \geq \Wc (z) > W_{i+1} \, \left(r_{i,i+1}\right) ^ {-h_{i,i+1}/2}\right\}$. The hysteresis width $h$ does not have to be identical for all boundaries, therefore we introduced the indices.

\subsection{Rational Instead of Integer Weight Ratios $r$}
\label{nonintegerWght}
Our roulette methods can already cope with more general weight ratios. $r$ has to be a rational number for the correlated roulette, while for the traditional Russian roulette and the deterministic roulette real $r$ would be possible as well. Splitting is more problematic. After splitting, the newly created markers should have equal weight because markers with less weight require just as much computational time as markers with higher weight, but contribute less to the result. This leads to an increased variance as compared to markers that are created with equal weights. 

We can still use non-integer weight ratios when we do not always split into the same number of markers. If we used uncorrelated random numbers to decide into how many markers to split, the total weight in the simulation would fluctuate strongly for long simulation times, just as with the Russian roulette. While the deterministic roulette would still avoid those fluctuations, we can eliminate them independently of the used roulette scheme by using correlated random numbers. With the deterministic roulette we can also avoid fluctuations of the higher moments by using an independent correlated random number generator for each high-weight marker separately. The details are presented elsewhere \cite{schuster2019MscThesis}.

\subsection{Choice of Parameters}
\label{parameterChoice}

First one has to decide on $\Wc$: The highest resolution for low densities is achieved by setting $a=1$ in equation \eqref{eq:combinedWghtFct}. But by setting an intermediate value for $a$ we can resolve both high and low densities well as we will explore in section \ref{flow}. To finish the definition of our weight regions we have to set the weight ratios between the regions. The exact value of this quantity is not critical. Smaller values give weight regions that more accurately follow the density while increasing the overhead due to reweighting.

Next we have to decide on the time that should elapse between two reweightings of all markers. We want to let the markers mix across a region before reweighting again instead of reweighting immediately, both to avoid detrimental effects due to consecutive reweightings and to reduce the computational overhead. Again, the exact value is not critical. An estimate of the mixing time for diffusive processes across a length $l$ can be found with 
\begin{align}
    t_\mathrm{mix} = \frac{l^2}{D},
\end{align}
with $D$ being the diffusion coeffient.
The width of a region $l$ and $D$ can change over the simulation domain and during runtime, and we have to estimate the typical value. If the mixing times differ strongly we can base $l$ on the smaller values of $t_\mathrm{mix}$, and use an additional hysteresis for the larger ones. A reasonable value for $h$ from section \ref{hyst} that allows mixing across a region is 0.8.

Empirically the timestep for reweighting should be $\approx t_\mathrm{mix}/10$, but as we will see in section \ref{ascotSection} even deviations from this rule by almost an order of magnitude still give good results. If reweighting with this estimated frequency is still too large of an overhead one can reduce the weight ratio between regions to increase $t_\mathrm{mix}$.

Finally it can be advantageous or necessary to specify a weight below which markers are no longer split into smaller ones.

\section{Characterization and Performance of Novel Schemes}
\label{results}

We characterize the schemes by considering simple diffusion-advection equations in sections \ref{wiener} and \ref{flow}. We let the simulations reach steady state and use bins to calculate the phase space density for many, i.e. $10^6$, points in time. From this set of measurements for the same phase space densities one can then calculate the variance of the density measurement. 

After splitting, the markers decorrelate on the length scale of the bin size. To analyze the decorrelation process it is therefore necessary to use overlapping bins.

To be able to compare simulations with different numbers of markers we have to eliminate the dependency of the error on the number of markers $N$. For simulations utilizing the deterministic roulette, or ones without reweighting, we can make use of the error scaling as $\mathcal{O}\left(N^{-1/2}\right)$ common for Monte Carlo simulations because the markers are independent. This scaling also holds for the correlated roulette, as we have shown numerically in a previous work~\cite{schuster2019MscThesis}. Note that the same scaling appears in equation \eqref{eq:eRelNoRewei} in a slightly different setting.

We therefore multiply the determined variance of some quantity with $\left<N\right>$, the mean number of markers used in the simulation. To obtain a quantity for the relative error we additionally divide by the squared mean of the quantity, $\mu$:
\begin{align}
	\sigma ^2 _\text{norm} = \sigma ^2 \frac{\left<N\right>}{\mu ^2}.
\end{align}

In the following we consider three different cases. More cases can be found in Schuster's thesis~\cite{schuster2019MscThesis}.

\subsection{Wiener Process}
\label{wiener}
We start our analysis with the Wiener process, a simple diffusive process, which is described by
\begin{align}
\label{eq:wiener}
\partial _t f = \left(\partial_x ^2 + \partial_y ^2\right) f.
\end{align}
The computational domain is two-dimensional with $x \in [0,1]$ and $y \in [-5, 5]$, with a common diffusion coefficient for both dimensions and mirror boundary conditions. The analytical solution for the steady state is $f(x, y, t) = const$. 
In more than one dimension the correlated roulette can show detrimental effects, which is why we consider the 2D case. We choose the $y$ domain larger than the $x$ domain to amplify these detrimental effects. The bins for the density measurement have a size of $0.2 \times 0.2$ and are positioned at $y=0$. For showing the effect of multiple consecutive crossings we want to choose a small time step for the Euler-Maruyama scheme \cite{kloeden2013numerical} used to advance the markers. As a compromise with the required computational time we choose $\approx \SI{2.2e-4}{}$. The initial distribution of markers is already constant in phase space. We start the simulations with 1000 markers.

\begin{figure}
\centering
\includegraphics[width=0.5\textwidth]{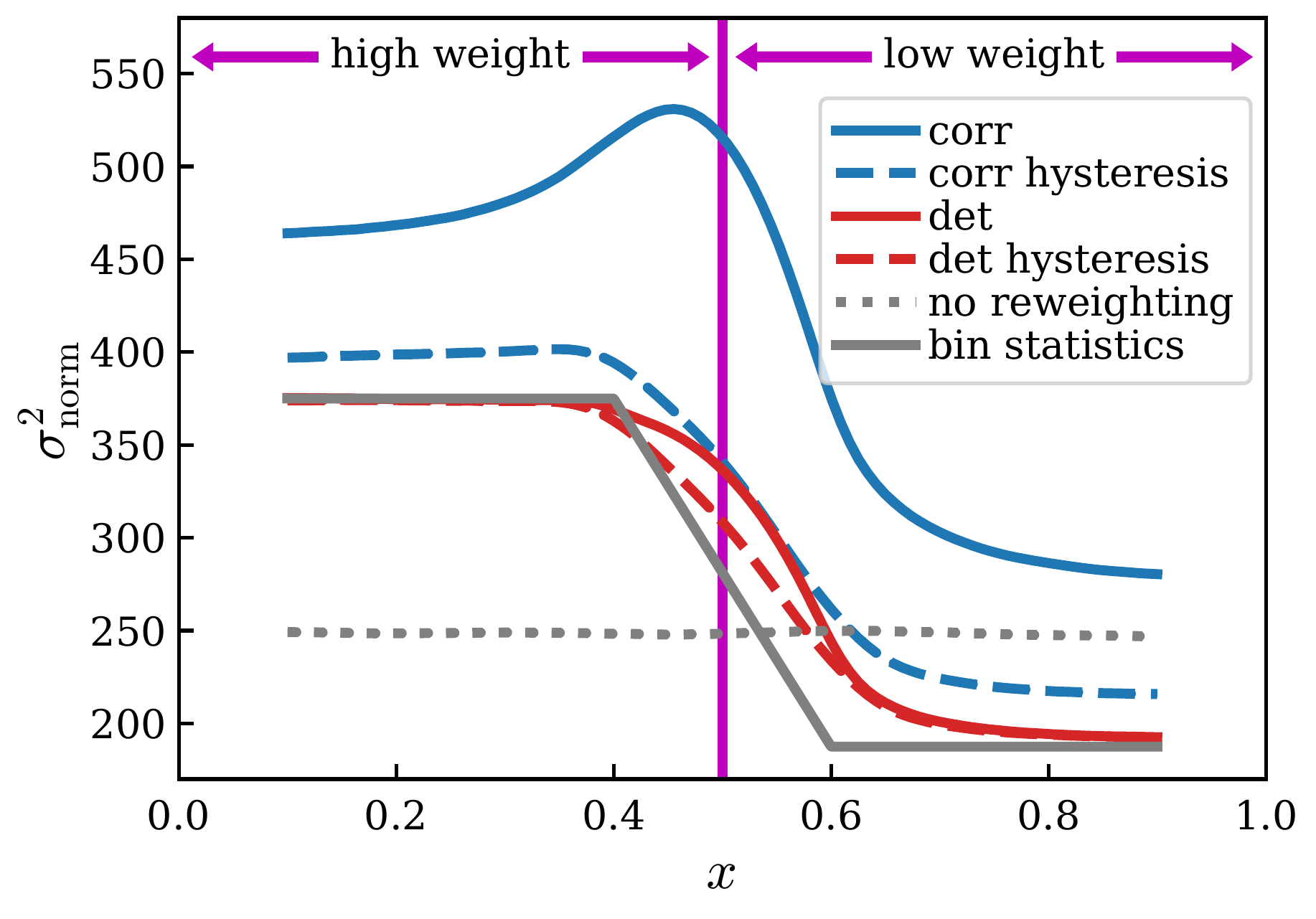}
\caption{The normalized variance of a binned density measurement in a two-dimensional domain. At $x<0.5$ we use markers with twice the weight as at $x>0.5$. We use markers of equal weight when we do not reweight. In the case presented here the correlated roulette performs worse than the deterministic roulette, especially when no hysteresis is used. This is not the case when the markers mix well in the $y$-direction (not shown here).}
\label{fig:wiener2d}
\end{figure}

\begin{table}
    \centering
    \begin{tabular}{l|c|r}
        Scheme & Hysteresis & \makecell{Relative mass error \\ using $1000$ markers} \\
        \hline
        no reweighting & - & \SI{0.0}{\percent}\\
        correlated & 0.0 & \SI{1.3}{\percent}\\
        correlated & 0.1 & \SI{0.9}{\percent}\\
        deterministic & 0.0 & \SI{2.2}{\percent}\\
        deterministic & 0.1 & \SI{1.8}{\percent}\\
    \end{tabular}
    \caption{The roulette schemes do not exactly conserve mass. In this table we show the relative mass error for the simulations in figure \ref{fig:wiener2d}. The correlated roulette maintains the mass more accurate than the deterministic roulette. A hysteresis reduces fluctuations not only for the binned density but also for the mass.  While the question on whether these fluctuations are problematic depends on the specific use case, we would generally negate this: Simulations tend to use orders by magnitude more markers than the 1000 we use for this example, and the error generally shrinks as $\mathcal{O}\left(N^{-1/2}\right)$ when using more markers. When we use the correlated roulette the error even scales as $\mathcal{O}\left({N^{-1}}\right)$ instead (section \ref{corrRoulette}). Note that these fluctuations do not grow with time.}
    \label{tab:massErrorWiener}
\end{table}

In Figure \ref{fig:wiener2d} we see the numerically determined $\sigma ^2 _\text{norm}$ for this setting. With markers of equal weight (gray, dotted line) $\sigma ^2 _\text{norm} = 249$\footnote{Consider a single marker. The probability distribution for its position is uniform. The result of the histogram is given by $\mathds{1}_\mathrm{bin}(\vec{z}_\mathrm{marker}) / \int_\mathrm{bin} \dif \vec{z}$. The definitions of mean, variance and $\sigma^2_\mathrm{norm}$ can now be applied, and we obtain the normalized variance for the histogram for markers without (re)weighting.}. 
We can estimate $\sigma ^2 _\text{norm}$ analytically using bin statistics (gray, solid line).
For reweighting we separate the domain into two weight regions: for $x<0.5$ the markers have twice the weight as for $x>0.5$. When we use the correlated roulette without hysteresis (blue, solid line) the results are less accurate than without reweighting everywhere, and lie clearly above the prediction from bin statistics. At $x\approx 0.5$ we see an increase in variance due to fluctuations arising from reweighting. When we use a hysteresis region of 0.1 (blue, dashed line) the markers are split and deleted less often, and the accuracy is improved. $\sigma ^2 _\text{norm}$ still lies above the prediction from bin statistics.

With the deterministic roulette the markers in the high weight region are not influenced by reweighting, as discussed in Section \ref{detRoulette}. This manifests in Figure \ref{fig:wiener2d} as a good agreement between the result with the deterministic roulette and the prediction from bin statistics for $x<0.4$. At $x>0.4$ there are already markers from the low weight region inside the bin because the bin width is 0.2. When going towards larger $x$, $\sigma ^2 _\text{norm}$ decreases, but it lags behind the prediction from bin statistics because the markers need time to decorrelate. When introducing a hysteresis of length 0.1 (red, dashed line) we reduce fluctuations and smooth the transition between the weight regions. 

Finally we also investigate the effect of reweighting on the particle mass in the system. As shown in table \ref{tab:massErrorWiener} here the correlated roulette leads to smaller deviations. A hysteresis reduces the deviations, as for the binned density. We would expect that these fluctuations in the total mass are not problematic for most applications.

\subsection{Diffusion-Advection Equation}
\label{flow}
In Section \ref{wiener} we looked at a single boundary between two weight regions in detail. There, $f$ was constant. Reweighting is however designed to improve the accuracy in low-density regions. To obtain an equilibrium density that varies as a function of $x$ we add advection to equation \eqref{eq:wiener}. For simplicity we restrict ourselves to a 1D process:
\begin{align}
\label{eq:diffAdvConstCoeff}
\partial _t f = -v \partial _x f + D \partial _x ^2 f.
\end{align}
Together with mirror boundary conditions we obtain the steady state solution
\begin{align}
\label{eq:diffAdvSteadyState}
f(x) = f_0 \exp{\left(-s x\right)}
\end{align}
with
\begin{align}
s = -\frac{v}{D} \approx 13.3.
\end{align}
We simulate this equation in the domain $\left[0,1\right]$, using mirror boundary conditions. The time step is \SI{2.5e-5}{}. 

As described in Section \ref{weightRegions}, we place the boundaries between weight regions such that the marker density lies between constant marker density and the marker density for uniform markers.
At $a=0$ the markers have equal weight, and at $a=1$ the weight of the markers is proportional to $f$. We use the deterministic roulette because it is more reliable in multiple dimensions as show in figure \ref{fig:wiener2d}, but the result is very similar for the correlated roulette in this case.

\begin{figure}
\centering
 \includegraphics[width=0.5\textwidth]{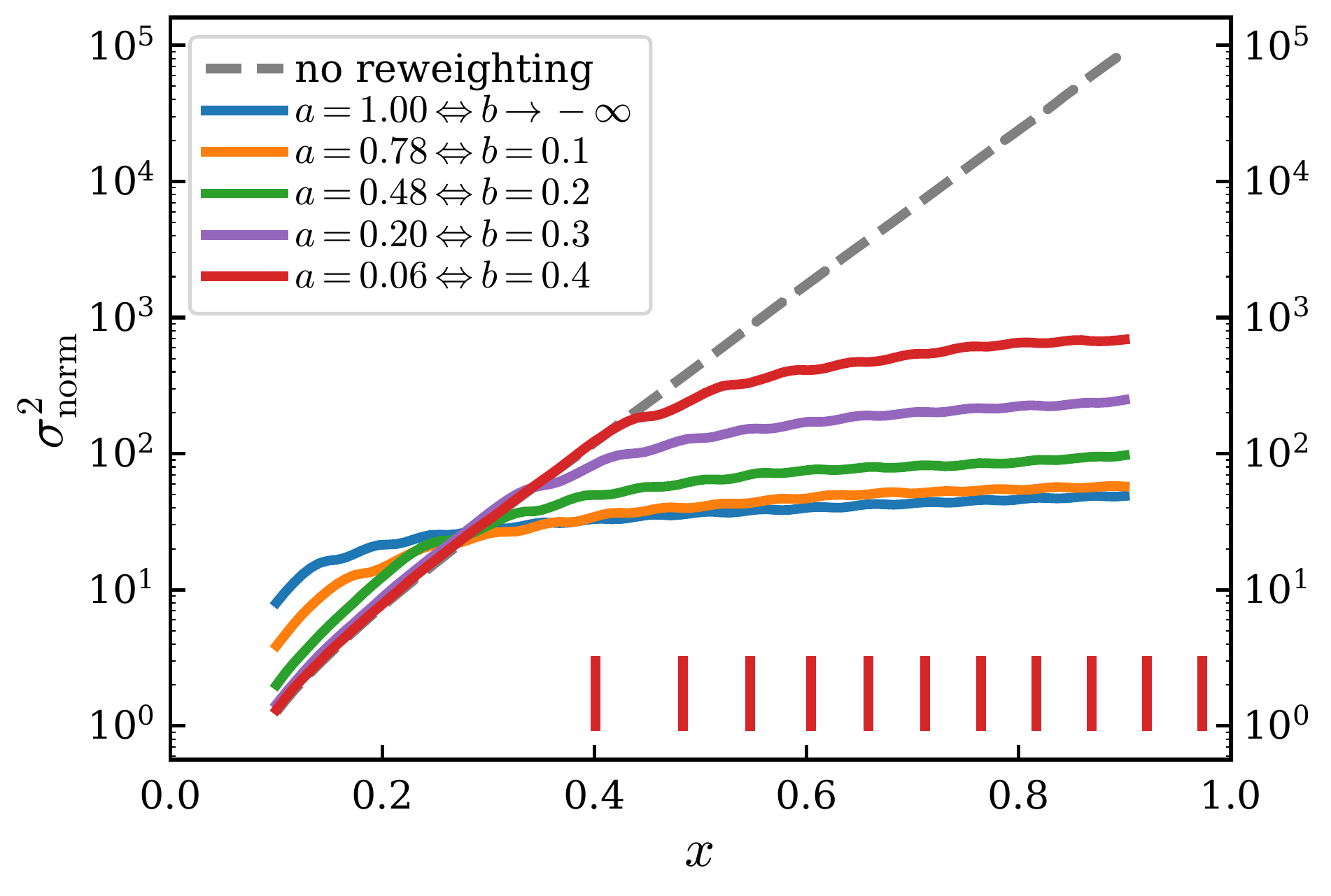}
 \caption{The normalized variance of density measurements at different positions $x$, shown for different $\Wc$ as defined by equation \eqref{eq:combinedWghtFct}. The boundaries between the weight regions for $b=0.4$ are shown as vertical red lines.}
\label{fig:interpWghtFctB04}
\end{figure}

\begin{table}
    \centering
    \begin{tabular}{l|r}
        $a$ & \makecell{Relative mass error \\ using $1000$ markers} \\
        \hline
        0 (no reweighting) & \SI{0.0}{\percent}\\
        1 & \SI{3.07}{\percent}\\
        0.78 & \SI{1.73}{\percent}\\
        0.48 & \SI{0.74}{\percent}\\
        0.20 & \SI{0.33}{\percent}\\
        0.06 & \SI{0.17}{\percent}\\
    \end{tabular}
    \caption{The relative mass error for the simulations in figure \ref{fig:interpWghtFctB04}. Regardless of the weighting of the particles the mass is always concentrated at small $x$. Smaller $a$ values, and therefore less aggressive reweighting at these higher densities, lead to smaller deviations. As argumented in table \ref{tab:massErrorWiener} we would expect that such fluctuations are unproblematic for most applications. Note that these fluctuations do not grow with time.}
    \label{tab:massErrorFlow}
\end{table}

The normalized variances $\sigma ^2 _\text{norm}$ of the simulations are shown in Figure \ref{fig:interpWghtFctB04}. The boundaries between the weight regions for $a=0.06$ are shown as vertical red lines. The density $f$ follows equation \eqref{eq:diffAdvConstCoeff} and is exponentially declining on a length scale of $v/D \approx 13.3$. Without reweighting, the variance is roughly inversely proportional to the density, leading to inaccurate results at $x=0.9$. By reweighting we can reduce the variance at large $x$ by more than three orders of magnitude. Simultaneously the variance at $x=0.1$ is increased by up to a factor of 8. With the parameter $a$ we can choose which regions we want to resolve with which accuracy. As an example, suppose we want to increase the accuracy in the low density region without compromising in the high density region, compared to the case without reweighting. By choosing $a=0.06$ we could reduce the variance at $x=0.9$ by over two orders of magnitude while increasing the variance at $x=0.1$ by merely \SI{3.4}{\percent}. 

There is a descriptive way of understanding how such large gains in accuracy in the low density tail can be achieved by sacrificing comparably little accuracy in the high density bulk. At high densities there are orders of magnitude more markers than at low densities. Moving a small fraction, e.g. a few percent, of the markers at high densities to low densities has only a small impact on the high density region. But it greatly increases the number of markers, and therefore the accuracy, in the low density region.

We show the fluctuations of the total mass that arise from reweighting in table \ref{tab:massErrorFlow}. If we decrease $a$ we reweight the high density areas less agressively, leading to fewer fluctuations. As high density areas contribute strongly to the total mass we therefore decrease fluctuations thereof by decreasing $a$. As before we suspect most applications of the schemes are not hindered by the mass fluctuations.

\subsection{ASCOT-RFOF}
\label{ascotSection}
As last test case we return to the physical problem of ICRH. The orbit-following code ASCOT \cite{hirvijoki2014ascot} calculates the distribution function of particles in tokamaks or stellarators. Together with the code library RFOF \cite{johnson2011library} ASCOT can be used to model ICRH. We implemented reweighting with both the deterministic and the correlated roulette in ASCOT. The weight regions are adjusted automatically, as described in Section \ref{weightRegions}. We use weights proportional to the distribution function, because the distribution function changes drastically during heating with RF waves. Reweighting is not yet parallelized even though the schemes are suited for parallelization. 

\begin{table}
    \centering
    \begin{tabular}{l|r}
        Quantity & Value \\
        \hline
        central electron temperature & \SI{7.2}{\kilo\electronvolt}\\
        central ion temperature & \SI{2.1}{\kilo\electronvolt}\\
        central electron density & \SI{4.3e19}{\per\cubic\meter}\\
        central deuteron density & \SI{1.7e19}{\per\cubic\meter}\\
        central proton density & \SI{0.36e19}{\per\cubic\meter}\\
        central nitrogen density & \SI{0.07e19}{\per\cubic\meter}\\
        \hline
        toroidal magnetic field & \SI{-2.4}{\tesla}\\
        plasma current & \SI{700}{\kilo\ampere}\\
        neutral beam heating & \SI{2.6}{\mega\watt}\\
        ion cyclotron heating & \SI{3.6}{\mega\watt}\\
        electron cyclotron heating & \SI{1.3}{\mega\watt}\\
        \hline
        simulated time & \SI{0.1}{\second}\\
        $a$ in equation \eqref{eq:combinedWghtFct} & 1\\
        time between reweighting & \SI{150}{\micro\second}\\
        weight ratio between regions& 1:3 \\
        hysteresis width $h$& 0.8\\
        number of regions & 10\\

    \end{tabular}
    \caption{Key experimental quantities of ASDEX Upgrade discharge \#33147 at $t=\SI{1}{\second}$ and parameters for the ASCOT simulation thereof. More details of the simulation setup can be found in Sipil\"a et.al. \cite{sipila2018monte}.}
    \label{tab:ascotParms}
\end{table}

\begin{figure}
\centering
\includegraphics[width=0.5\textwidth]{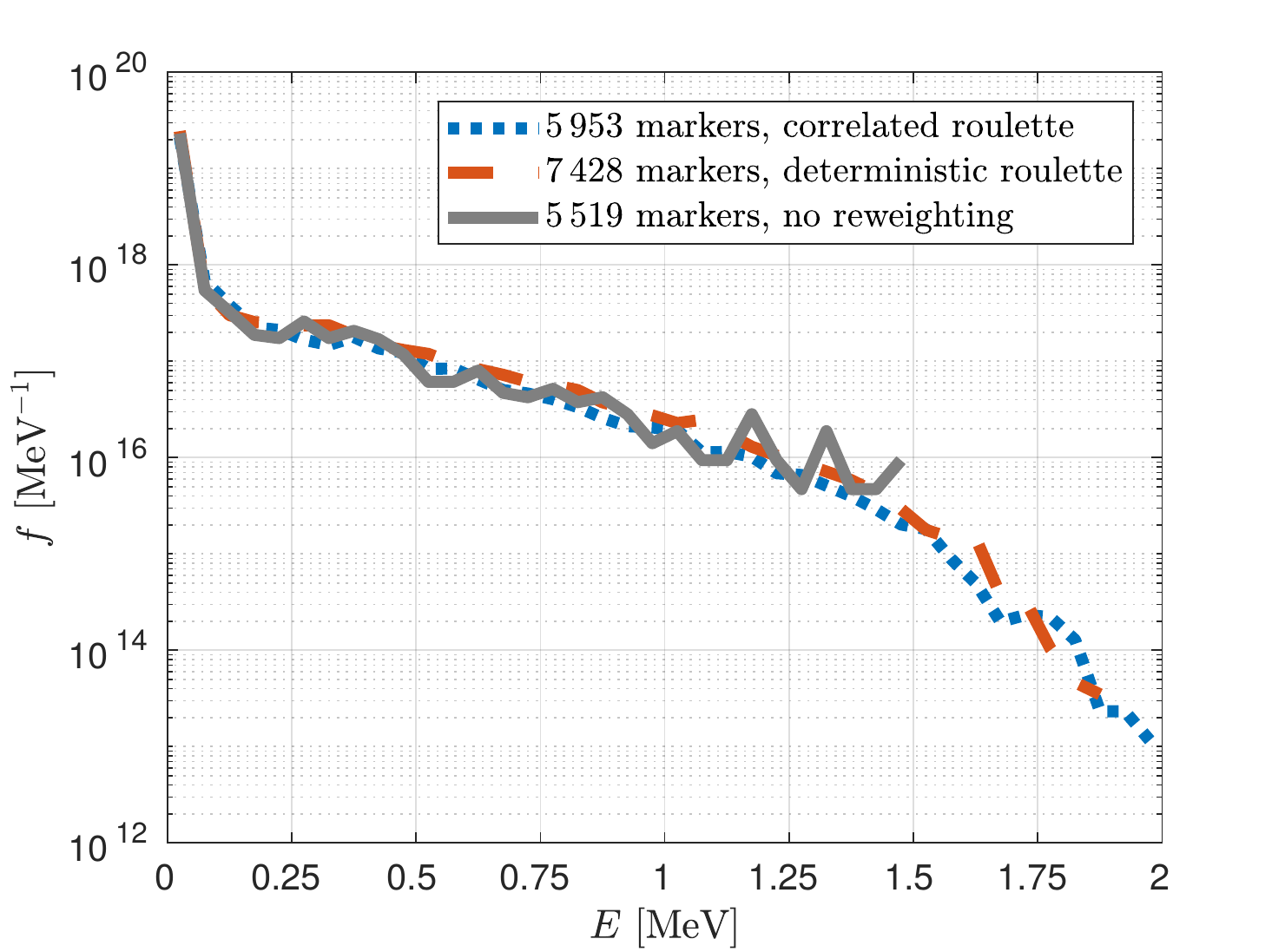}
\caption{Hydrogen distribution functions calculated with ASCOT-RFOF. The setup of the simulations, which model the ASDEX Upgrade discharge \#33147, is based on Sipil\"a et.al. \cite{sipila2018monte}. With reweighting we can resolve lower densities and reduce the Monte Carlo noise. The average absorbed heating power for the correlated, deterministic and equally weighted cases are \SI{2.24}{\mega\watt}, \SI{2.28}{\mega\watt} and \SI{2.29}{\mega\watt}.  The simulations were performed on the DRACO cluster at IPP Garching.}
\label{fig:ascotDistFct}
\end{figure}

\begin{figure}
\centering
\includegraphics[width=0.5\textwidth]{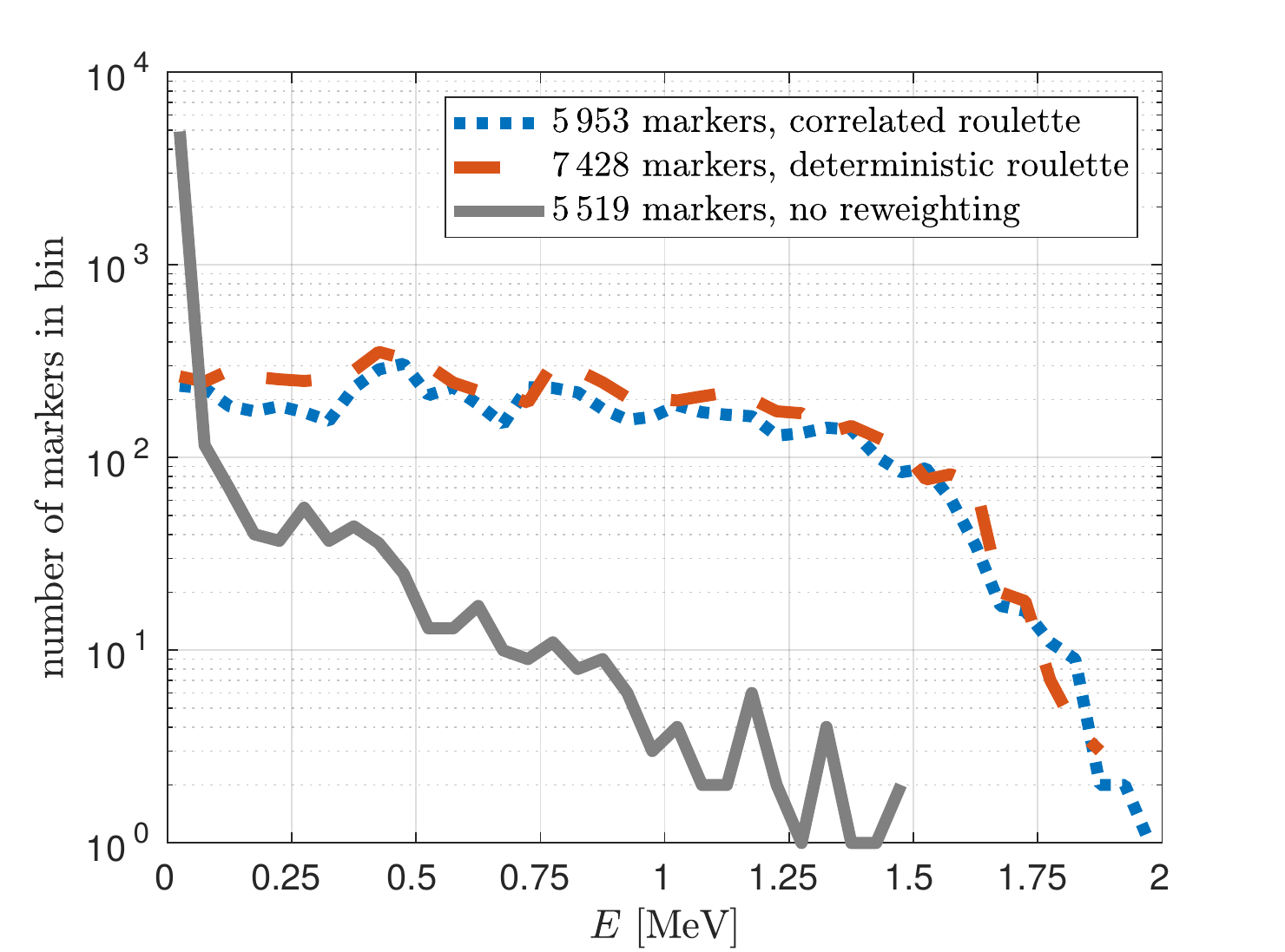}
\caption{The number of markers in each bin, with a width of \SI{50}{\kilo\electronvolt} each. As specified in table \ref{tab:ascotParms} we use 10 weight regions with a weight ratio of 1/3 each. By choosing $a=1$ in equation \eqref{eq:combinedWghtFct} we specify constant marker density until the density drops by a factor of $(1/3)^{10-1}\approx \SI{5e-5}{}$. From figure \ref{fig:ascotDistFct} we can determine that the distribution drops below this factor at $E \approx \SI{1.6}{\mega\electronvolt}$. We show here that the marker density indeed is approximately constant until $E \approx \SI{1.6}{\mega\electronvolt}$ when we reweight the markers with the presented methods (red and blue lines). At higher energies, corresponding to lower densities, we do not reweight any further leading to a drop of marker density. If we use markers of equal weight on the other hand (grey line) most markers are concentrated at low energies below \SI{100}{\kilo\electronvolt}. As we discuss in figure \ref{fig:ascotNoise} the higher marker density reduces the Monte Carlo noise by orders of magnitude at high energies.}
\label{fig:ascotNumMarkers}
\end{figure}

ASCOT-RFOF is initialized with a wave amplitude which determines the absorbed heating power. If we keep this wave amplitude constant the absorbed power is initially only $\approx \SI{0.05}{\mega\watt}$, but increases to an average $\approx \SI{2.3}{\mega\watt}$ during \SI{0.1}{\second} as the distribution function evolves. Usually one adjusts the wave amplitude during the simulation such that the absorbed power matches the desired value throughout the simulation.
This feedback constitutes a nonlinearity in the equation of motion of the Monte Carlo markers. A certain number of markers is required such that the absorbed power can be determined accurately. Because reweighting is not yet parallelized we can only use a limited number of markers, and we cannot readjust the RF wave amplitude. Instead the wave amplitude is kept constant throughout the simulation, resulting in a different heating power compared to the experiment. The difference in heating power to the experiment does not hinder comparing the numerical methods. We also do not replenish wall losses to simplify the implementation. The simulation results can therefore not be compared to the experiment, but we can nevertheless compare simulations with and without reweighting.

The hydrogen distribution function $f$ is calculated with ASCOT-RFOF, based on the setup from Sipil\"a \cite{sipila2018monte} for ASDEX Upgrade discharge \#33147. Parameters for the modelled discharge and the simulations themself are given in table \ref{tab:ascotParms}.
We follow the procedure from section \ref{parameterChoice} for choosing the tabulated values. We place our regions such that we put maximal emphasis on the low-density tail of the distribution function by setting $a=1$ in equation \eqref{eq:combinedWghtFct}. When choosing the timestep for reweighting we are facing vastly varying diffusivities as a function of the energy coordinate. For thermal energies we would ideally choose a reweighting timestep of $\approx \SI{20}{\micro\second}$, for higher energies the mixing time would suggest a much larger value. To avoid overhead we select a comparably large reweighting timestep of $\approx \SI{150}{\micro\second}$, and use a hysteresis to allow the markers to mix also at higher energies.

We compare simulations without reweighting and with our new weighting schemes in Figure \ref{fig:ascotDistFct}. All three distribution functions agree well. With reweighting we can resolve energies that are orders of magnitude lower than the ones resolvable with uniform weights ('no reweighting' in figure \ref{fig:ascotDistFct}). Additionally the distribution functions obtained with reweighting show less noise. These advantages stem from the tailored marker density, which is shown in figure \ref{fig:ascotNumMarkers}. 

\begin{figure}
\centering
\includegraphics[width=0.5\textwidth]{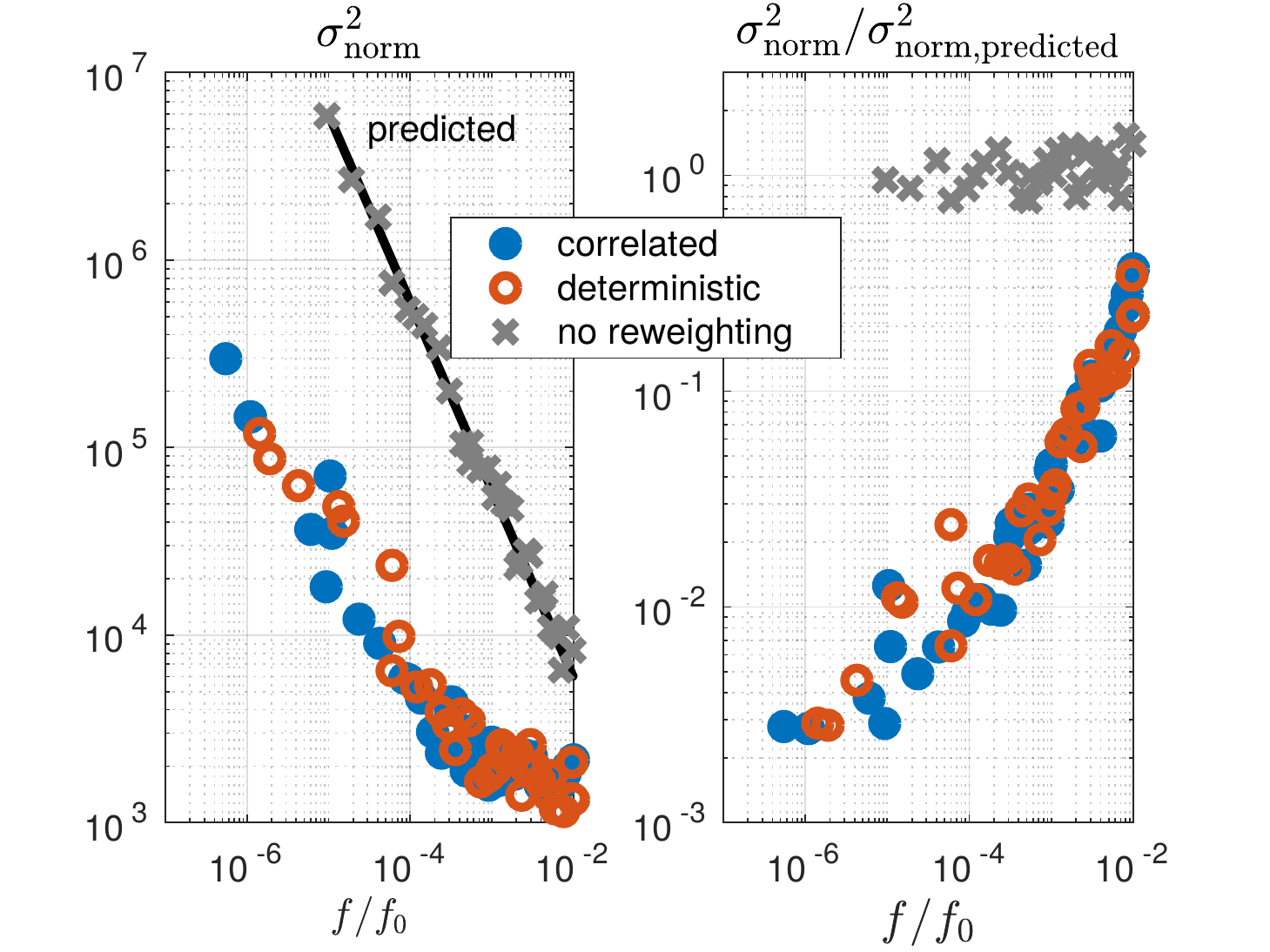}
\caption{On the left we show $\sigma_\mathrm{norm}$ as described in \ref{ascotSection} as a function of the distribution function. The distribution function is normalized to the density of thermal ions. When going towards lower densities the normalized variance increases more slowly with reweighting than without it. Both reweighting schemes perform similarly. Bin statistics can predict the accuracy of the simulation without reweighting, the prediction is shown as black line. We divide all points by the prediction by bin statistics for the case without reweighting, and show the result on the right to be able to judge the improvement from reweighting directly. At $f/f_0 = 10^{-6}$, corresponding to ions of $\approx \SI{2}{\mega\electronvolt}$, the variance due to noise is reduced by a factor of $\approx 350$. The data without reweighting is from Sipil\"a \cite{sipila2018monte}.}
\label{fig:ascotNoise}
\end{figure}

To quantify the Monte Carlo noise we can not use the same approach of the simpler test cases in Sections \ref{wiener} and \ref{flow} because the simulation is too expensive to be run for long enough times. Bin statistics is also not applicable because the markers are correlated, and therefore no Poisson distribution for the number of markers in the bins can be assumed. Instead, we separate each bin from the histogram shown in Figure \ref{fig:ascotDistFct} into 50 smaller bins. We assume that $f$ changes only little over \SI{50}{\kilo\electronvolt} and interpret the values from the 50 smaller bins as samples of the same physical $f$. We can thus calculate the variance $\sigma ^2$, and therefore $\sigma ^2_\mathrm{norm}$, for each histogram bin. This yields only an estimate for the variance due to noise: We are unable to detect offsets that span over \SI{50}{\kilo\electronvolt}, and any physical change of the distribution function over \SI{50}{\kilo\electronvolt} is interpreted as noise.

In Figure \ref{fig:ascotNoise} we show $\sigma ^2_\mathrm{norm}$ as a function of $f$. At high densities the distribution function varies quickly, violating our assumption of constant density in each bin. It is therefore not possible to reconstruct the variance for $f/f_0>10^{-2}$, where $f_0 = f(E\myeq0)$. With reweighting the results are more accurate than without reweighting already at $f/f_0>10^{-2}$. When going towards lower densities the normalized variance increases more slowly with reweighting than without it. Both new reweighting schemes perform identically. We can predict the normalized variance of the result without reweighting using bin statistics (prediction is shown as black line). We then divide all data points by the prediction to compare the performance more directly, this is shown on the right. At $f/f_0 = 10^{-6}$, corresponding to ions of $\approx \SI{2}{\mega\electronvolt}$, the variance due to Monte Carlo noise is reduced by a factor of $\approx 350$. Observe that due to noisy flux there are correlations between bins, meaning noise is not the only source of the variance of the result.

When interested in ions with an energy of \SI{2}{\mega\electronvolt} we can reduce the computational time by a factor of 350, provided enough markers are used to renormalize the RF wave field.

\section{Conclusion}

When using splitting and roulette schemes one has to define regions in the simulation domain and boundaries between them. Every region is associated with the weight of the markers therein. We presented a method to automatically define these regions, and even adapt them during the simulation to account for changing requirements due to the evolving distribution function. We developed two new roulette schemes because the traditional Russian Roulette \cite[p. 99]{hammersley2013MCmethods} leads to fluctuations that accumulate during the simulation. The new roulette methods possess two important properties:
\begin{itemize}
	\item Approximate conservation of mass for the correlated roulette, and of all moments of $f$ for the deterministic roulette.
	\item Nonexistence of a bias: The mean result of simulations is the same as without reweighting. This is proven for a special case in \ref{unalteredMean} and shown numerically elsewhere \cite{schuster2019MscThesis}.
\end{itemize}

The schemes are accompanied by various methods to improve accuracy (Section \ref{hyst}) and that simplify use and implementation of the schemes, for example an algorithm for efficient and easy initialization of markers, and a manual for choosing parameters \cite[4.4]{schuster2019MscThesis}. 

We verified the new schemes with simple models that allow for an analytical solution, and showed that they are capable of reducing the variance due to Monte Carlo noise by orders of magnitude. This is also the case with realistic applications, as in the case of fast-ion tails generate with ICRF heating in fusion plasmas.

Simulations of ICRF heating are challenging because coefficients of the governing Fokker-Planck equation depend on time and nonlocally on the solution. Furthermore source terms can be present from neutral beam injected particles or fusion reactions, which require that markers are added during runtime. The algorithms presented in this paper however can cope with these requirements. Their implementation in existing codes is relatively straight forward, since the schemes do not alter the equations of motion of the markers. Finally, the reweighting schemes can be parallelized and require minimal communication between threads.

\section*{Acknowledgments}

We would like to thank Jakob Ameres, Roman Hatzky and Omar Maj for fruitful discussions. This work has been carried out within the framework of the EUROfusion Consortium and has received funding from the Euratom research and training programme 2014-2018 and 2019-2020 under grant agreement No 633053. The views and opinions expressed herein do not necessarily reflect those of the European Commission.

\bibliographystyle{elsarticle-num}
\bibliography{bibliography.bib}

\appendix
\onecolumn
\section{Unaltered Mean Result of Simulation}
\label{unalteredMean}
In this section we show for a special case that the deterministic roulette does not alter the mean result of the simulation. The case that we consider is restricted as follows:
\begin{itemize}
	\item There is only one boundary.
	\item There is no hysteresis.
	\item The weight in the high weight region is twice the weight in the low weight region.
\end{itemize}

\subsection{Problem Description and Definitions}
Markers are at discrete positions in a 1D domain, and advance on discrete time steps. The change from continuous to discrete space and time is no restriction as the numerical schemes are used on computers, which require discretization anyways. The positions with index $i<0$ are the high weight region, the positions with $i\geq0$ are the low weight region with half the weight. We will call the marker that originated in the high weight region 'blue', and the markers that are created by the blue marker 'green'. 

We only consider a single blue marker. This is however no restriction as the blue marker, and the green markers it creates, is completely independent and therefore uncorrelated from any other blue or green markers.

The green and blue markers travel from a position $p$ to other positions $q$ with probabilities $G(p,q)$ and $B(p,q)$, respectively. As there is no restriction to $B$ and $G$, the marker position can be interpreted as a label for grid points in a multidimensional domain. This proof therefore automatically also holds for multidimensional problems. When a green marker jumps to the high weight region it is deleted. When the blue marker jumps from the high weight region to the low weight region it creates a green marker at the target position. When a green marker has been deleted, we denote its state as $D\notin\N_0$.

The set of states the markers can be in is given by
\begin{align}
	X = \left\{\left(b, \vec{g}\right) | b \in \Z, g \in \left(\N_0 \cup \{D\}\right)^n, n\in\N_0\right\},
\end{align}
i.e. there are $n$ green markers, and every marker has a position. The number of elements in $\vec{g}$ is $n$. The probability space for these states is given by \footnote{Since we are only interested in finite times, only a finite number of green markers can be generated. Therefore a probability distribution on $X$ can be seen as an element of the space of finite sequences, indexed by $X$.}
\begin{align}
	C_{00}(X) = \mathrm{span}\left(\ket{b, \g}, x \in X\right). 
\end{align}
A (properly normalized) state $\ket{\Psi} \in C_{00}$ is called the stochastic vector, and can be interpreted as probability distribution on $X$: If $\ket{\Psi} \in X$ we are certain what the state of the system is. If $\ket{\Psi} \notin X$ the system could be in different states with some probabilities. The probabilities are the coefficients for the basis vectors of $C_{00}$.
The transition probability between states is given by
\begin{align}
	m\left(x\to y\right) = P\left(\mathrm{state} = y \mathrm{~at~time~} t+1 | \mathrm{state} = x \mathrm{~at~time~} t\right),
\end{align}
which is the $x,y$ element of the corresponding stochastic matrix $M$. A state at time $t+1$ can be obtained from the state at time $t$:
\begin{align}
	\ket{\Psi}_{t+1} = M \ket{\Psi}_t
\end{align}
Our main quantity of interest will be the number of green markers at the position $p$:
\begin{align}
	\label{eq:proofFpDef}
	F_p\left(\ket{b,\g}\right) = \#\left\{i | \left(\g\right)_i = p\right\} = \sum_{k=1}^\len{\g} \delta_{g_k,p}.
\end{align}
$\len{\g}$ is the number of entries in $\g$, which is the number of green markers that have been created. Similarly, the number of blue markers at the position $p$ is given by
\begin{align}
	H_p\left(\ket{b,\g}\right) = \delta_{b,p}.
\end{align}
We expand the definition sets of $F_p$ and $H_p$ by linearity to the space $C_{00}\left(x\right)$. This expansion by linearity is equivalent to calculating the mean density of markers as position $p$:
\begin{align}
	F_p\left(\ket{\Psi}\right) = F_p\left(\sum_i p_i \ket{b,\g}\right) = \sum_i p_i F_p\left(\ket{b_i,\g_i}\right) = \mean{F_p}_{\left\{p_i\right\}},
\end{align}
the same holds for $H$.

\subsubsection{Goal}
The reconstructed probability density of the stochastic process is calculated by binning the marker positions, and multiplying the number of markers in each bin with the weight of the markers. The mean density is then
\begin{align}
	\mean{\rho} = w_g \mean{F_p} + w_b \mean{H_p},
\end{align}
where $w_g$ and $w_b$ are the weights of the green and blue markers. In the high weight region $F_p$ is 0, and $w_b = 1$. The same is true if there is no reweighting. We then have
\begin{align}
	\mean{\rho} = \mean{H_p}.
\end{align}
In the low weight region we have $w_g = w_b = 1/2$, 
\begin{align}
	\mean{\rho} = \frac{1}{2}\left(\mean{F_p} + \mean{H_p}\right).
\end{align}
The proof is finished when we find that $\mean{\rho}$ is the same with and without reweighting. In other words,
\begin{align}
	\frac{1}{2}\left(\mean{F_p} + \mean{H_p}\right) = \mean{\rho} = \mean{H_p},
\end{align}
hence
\begin{align}
	\mean{F_p}_t = F_p\left(\ket{\Psi}\right)_t = H_p\left(\ket{\Psi}\right)_t = \mean{H_p}_t \label{eq:proofToShow}
\end{align}
in the low weight region ($p\geq0$) for all times $t$. 

\subsubsection{Outline}
In \ref{proofDensitiesAreEnough} we first show that $\mean{F_p}$ at time $t+1$ can be calculated with $\mean{F_p}$ and $\mean{H_p}$ at time $t$, without requiring $\ket{\Psi}$ itself. After this we only have to use induction to finish the proof of preposition \eqref{eq:proofToShow}. This is done in \ref{proofInduction}.
\subsection{Evolution of the Densities}
\label{proofDensitiesAreEnough}
We start with $F$ at $t+1$:
\begin{align}
	F_p\left(M \ket{\Psi}\right) &= \sum_i p_i F_p\left(M \ket{b_i,\g_i}\right) \\
	&= \sum_i p_i F_p\left(\sum_{\left(b',\g'\right) \in X}m\left(\left(b_i,\g_i\right) \to \left(b',\g'\right)\right) \ket{b',\g'}\right) \\
	&= \sum_i p_i \sum_{\left(b',\g'\right) \in X}m\left(\left(b_i,\g_i\right) \to \left(b',\g'\right)\right) F_p\left(\ket{b',\g'}\right) \\
	&= \sum_i p_i \sum_{b'=-\infty}^\infty \sum_{n=0}^\infty \sum_{(\g')_1, \dots, (\g')_n \in \N_0 \cup D}   m\left(\left(b_i,\g_i\right) \to \left(b',\g'\right)\right) F_p\left(\ket{b',\g'}\right). \label{eq:proofFpExpanded}
\end{align}
\footnote{Note that the sums are actually finite since $M$ maps from finite sequences to finite sequences (only up to one marker can be created in each time step).}We will now decompose $m\left(x\to y\right)$ further. For this we distinguish three cases: 
\begin{description}
	\item [All markers in state $y$ exist already in state $x$:] All markers must move to the correct position, and no new marker is created.
	\item [In state $y$ exists one additional green marker:] All markers must move to the correct position, and the blue marker has to create the additional green marker. This additional green marker has to be at the correct position.
	\item [Any other initial state $x$:] The probability to advance to state $y$ is 0.
\end{description}
We can express this through the probabilities of the blue or green markers to move from position $p_1$ to $p_2$: $B\left(p_1\to p_2\right)$ and $G\left(p_1\to p_2\right)$.
\begin{align}
	m\left(\left(b_i,\g_i\right) \to \left(b',\g'\right)\right) = B\left(b_i, b'\right) G\left((\g_i)_1, (\g')_1\right) \dots G\left((\g_i)_\len{\g_i}, (\g')_\len{\g_i}\right) \begin{cases} \delta_{\len{\g_i}+1, \len{\g'}} \delta_{\g'_\len{\g'},b'}& \mbox{if } b_i < 0 \mathrm{~and~} b' \geq 0 \\
	\delta_{\len{\g_i},\len{\g'}} & \mbox{else }\end{cases} \\
	= B\left(b_i, b'\right) G\left((\g_i)_1, \g'_1\right) \dots G\left((\g_i)_\len{\g_i}, \g'_\len{\g_i}\right) \bigcdot \left[\underbrace{\delta_{\len{\g_i},\len{\g'}}}_\mathrm{I} + \begin{cases} \underbrace{\delta_{\len{\g_i}+1,\len{\g'}} \delta_{(\g')_n,b'}}_\mathrm{II} - \underbrace{\delta_{\len{\g_i},\len{\g'}}}_\mathrm{III}& \mbox{if } b_i < 0 \mathrm{~and~} b' \geq 0 \\
	0 & \mbox{else }\end{cases}\right] \label{eq:proofTransitionProb}
\end{align}
Now we can insert equation \eqref{eq:proofTransitionProb} and the sum representation of equation \eqref{eq:proofFpDef} into equation \eqref{eq:proofFpExpanded}. Then equation \eqref{eq:proofFpExpanded} will decompose into three terms $T_\mathrm{I}$, $T_\mathrm{II}$ and $T_\mathrm{III}$ corresponding to the terms I, II and III in equation \eqref{eq:proofTransitionProb}

\begin{align}
	F_p\left(M \ket{\Psi}\right) = T_\mathrm{I} + T_\mathrm{II} - T_\mathrm{III}. \label{eg:proofDecomposedFp}
\end{align}
We will treat the three terms $T_\mathrm{I}$, $T_\mathrm{II}$ and $T_\mathrm{III}$ separately. We start with $T_\mathrm{I}$:
\begin{align}
	T_\mathrm{I} = \sum_i p_i \sum_{b'=-\infty}^\infty \sum_{n=0}^\infty \sum_{(g')_1, \dots, (g')_n \in \N_0 \cup D} B\left(b_i, b'\right) G\left((g_i)_1, g'_1\right) \dots G\left((g_i)_\len{\g_i}, g'_\len{\g_i}\right) \delta_{\len{\g_i},\len{\g'}} \sum_{k=1}^\len{\g'} \delta_{g'_k,p}.
\end{align}
We can immediately evaluate the sums over $b'$ and $n$. The probability of the blue marker moving anywhere is 1. Note that for the term I $\len{\g_i} = \len{\g'}$ since there is no new marker generated. Because $\len{\g'} = n$, $\delta_{\len{\g_i},\len{\g'}}$ will be 1 only for $\len{\g'} = n$, what is used to resolve the sum over $n$.
\begin{align}
	T_\mathrm{I} &= \sum_i p_i \sum_{(\g')_1, \dots, (\g')_\len{\g_i} \in \N_0 \cup D} G\left((\g_i)_1, (\g')_1\right) \dots G\left((g_i)_\len{\g_i}, (\g')_\len{\g_i}\right) \sum_{k=1}^\len{\g'} \delta_{(\g')_k,p} \\
	&=\sum_i p_i \sum_{k=1}^\len{\g_i} \sum_{(\g')_1, \dots, (\g')_\len{\g_i} \in \N_0 \cup D} G\left((\g_i)_1, (\g')_1\right) \dots G\left((\g_i)_\len{\g_i}, (\g')_\len{\g_i}\right) \delta_{(\g')_k,p} \\
	&= \sum_i p_i \sum_{k=1}^\len{\g_i} \sum_{(\g')_1} G\left((\g_i)_1, (\g')_1\right) \dots \sum_{(\g')_{k-1}} G\left((\g_i)_{k-1}, (\g')_{k-1}\right) \; G\left((\g_i)_k, p\right) \; \sum_{(\g')_{k+1}} G\left((\g_i)_{k+1}, \g'_{k+1}\right) \dots \sum_{(\g')_\len{\g_i}} G\left((\g_i)_\len{\g'}, (\g')_\len{\g_i}\right).
\end{align}
The sums over the positions of the markers that are not the $k$-th marker all yield 1: The probability for the marker to either survive or be moved to the 'parking position' $D$ is 1. 
\begin{align}
	T_\mathrm{I} &= \sum_i p_i \sum_{k=1}^\len{\g_i} G\left((g_i)_k, p\right).
\end{align}
We insert unity, $f\left((\g_i)_{k}\right) = \sum_{q=0}^\infty \delta_{(\g_i)_{k}, q} f\left(q\right)$. It is not necessary to include $q=D$ because a marker that was deleted and is therefore at $D$ can never leave $D$ again. We recognize the definition of $F$:
\begin{align}
	T_\mathrm{I} &= \sum_i p_i \sum_{k=1}^\len{\g_i} \sum_{q=0}^\infty \delta_{(g_i)_{k}, q} G\left(q, p\right) = \sum_i p_i \sum_{q=0}^\infty G\left(q, p\right) \sum_{k=1}^\len{\g_i} \delta_{(g_i)_{k}, q} = \sum_i p_i \sum_{q=0}^\infty G\left(q, p\right) F_q\left(\ket{b_i, \g_i}\right) = \sum_{q=0}^\infty G\left(q, p\right) F_q\left(\ket{\Psi}\right), \label{eq:proofTI}
\end{align}
where we used the linearity of $F$ in the last step. We can immediately interpret this result: if there was no blue marker, the term I would be the only term. Then the mean green marker density only depends on the mean marker densities of the green markers in the previous time step because the green markers evolve independently. 

We will now perform the same calculations with terms II and III. To satisfy the case distinction we restrict the sum over $b'$, and use the Iverson bracket, with $P$ being a logical expression: 
\begin{align}
\left[P\right] = \begin{cases}1 & \mbox{if } P \mathrm{~is~true} \\ 0 & \mbox{else } \end{cases}
\end{align}
\begin{align}
	T_\mathrm{II} = \sum_i p_i \sum_{b'=0}^\infty \left[b_i<0\right] \sum_{n=0}^\infty \sum_{(\g')_1, \dots, (g')_n \in \N_0 \cup D} B\left(b_i, b'\right) G\left((\g_i)_1, (\g')_1\right) \dots G\left((\g_i)_\len{\g_i}, (\g')_\len{\g_i}\right) \delta_{\len{\g'},\len{\g_i}+1} \delta_{(\g')_\len{\g'},b'} \sum_{k=1}^\len{\g'} \delta_{(\g')_k,p}.
\end{align}
Other than for $T_\mathrm{I}$ we have $\len{\g'} = \len{\g_i} + 1$ due to the marker we create, and we cannot resolve the $b'$ sum. 
\begin{align}
	T_\mathrm{II} &= \sum_i p_i \sum_{b'=0}^\infty \left[b_i<0\right] \sum_{(\g')_1, \dots, (\g')_{\len{\g_i}+1}} B\left(b_i, b'\right) G\left((\g_i)_1, (\g')_1\right) \dots G\left((\g_i)_\len{\g_i}, (\g')_\len{\g_i}\right) \delta_{(\g)'_{\len{\g_i}+1},b'} \sum_{k=1}^\len{\g'} \delta_{(\g')_k,p} \\
	&= \sum_i p_i \sum_{b'=0}^\infty \left[b_i<0\right] \sum_{(\g')_1, \dots, (\g')_{\len{\g_i}}} B\left(b_i, b'\right) G\left((\g_i)_1, (\g')_1\right) \dots G\left((\g_i)_\len{\g_i}, (\g')_\len{\g_i}\right) \sum_{(\g')_{\len{\g_i}+1}} \delta_{(\g)'_{\len{\g_i}+1},b'} \left[\delta_{(\g')_{\len{\g_i}+1},p} + \sum_{k=1}^{\len{\g'}-1} \delta_{(\g')_k,p} \right] \\
	&= \sum_i p_i \sum_{b'=0}^\infty \left[b_i<0\right] \sum_{(\g')_1, \dots, (\g')_{\len{\g_i}}} B\left(b_i, b'\right) G\left((\g_i)_1, (\g')_1\right) \dots G\left((\g_i)_\len{\g_i}, (\g')_\len{\g_i}\right) \left[\delta_{b',p} + \sum_{k=1}^\len{\g_i} \delta_{(\g')_k,p} \right]
\end{align}

For $T_\mathrm{III}$ we get
\begin{align}
	T_\mathrm{III} &= \sum_i p_i \sum_{b'=0}^\infty \left[b_i<0\right] \sum_{n=0}^\infty \sum_{(\g')_1, \dots, (\g')_\len{\g'} \in \N_0 \cup D} B\left(b_i, b'\right) G\left((\g_i)_1, (\g')_1\right) \dots G\left((\g_i)_\len{\g_i}, (\g')_\len{\g_i}\right) \delta_{\len{\g'},\len{\g_i}} \sum_{k=1}^\len{\g'} \delta_{(\g')_k,p} \\
	&= \sum_i p_i \sum_{b'=0}^\infty \left[b_i<0\right] \sum_{(\g')_1, \dots, (\g')_\len{\g_i}} B\left(b_i, b'\right) G\left((\g_i)_1, (\g')_1\right) \dots G\left((\g_i)_\len{\g_i}, (\g')_\len{\g_i}\right) \sum_{k=1}^\len{\g_i} \delta_{(\g')_k,p}.
\end{align}

We now subtract terms $T_\mathrm{II}$ and $T_\mathrm{III}$. The rightmost sum is identical for both terms an cancels, only the additional $\delta_{b',p}$ from term $T_\mathrm{II}$ survives.
\begin{align}
	T_\mathrm{II} - T_\mathrm{III} &= \sum_i p_i \sum_{b'=0}^\infty \left[b_i<0\right] \sum_{(g')_1, \dots, (g')_{\len{\g_i}}} B\left(b_i, b'\right) G\left((g_i)_1, g'_1\right) \dots G\left((g_i)_\len{\g_i}, g'_\len{\g_i}\right) \delta_{b',p} \\
	&= \sum_i p_i \sum_{b'=0}^\infty \left[b_i<0\right] B\left(b_i, b'\right) \delta_{b',p} \\
	&= \sum_i p_i \left[b_i<0\right] B\left(b_i, p\right) \left[p\geq0\right]. \label{eq:proofSubstractedTIITIII}
\end{align}
Analogously to what we did for term $T_\mathrm{I}$, we insert unity to recover $H$:
\begin{align}
	\eqref{eq:proofSubstractedTIITIII} &= \sum_i p_i \sum_{q = -\infty} ^\infty \delta_{b_i,q} \isneg{q} B\left(q, p\right) \left[p\geq0\right] \\
	&= \sum_i p_i \sum_{q = -\infty} ^{-1} H_q\left(\ket{b_i, \g_i}\right) B\left(q, p\right) \left[p\geq0\right] \\
	&= \left[p\geq0\right] \sum_{q = -\infty} ^{-1} B\left(q, p\right) H_q\left(\ket{\Psi}\right). \label{eq:proofSubstractedTIITIIIDone}
\end{align}
In the last step we used the linearity of $H$. We can also interpret this result: The blue marker that enters the low weight region acts as a source for the green markers.

Inserting \eqref{eq:proofTI} and \eqref{eq:proofSubstractedTIITIII} into equation \eqref{eg:proofDecomposedFp} we finally obtain
\begin{align}
	F_p\left(M \ket{\Psi}\right) = T_\mathrm{I} + T_\mathrm{II} - T_\mathrm{III} = \sum_{q=0}^\infty G\left(q, p\right) F_q\left(\ket{\Psi}\right) + \left[p\geq0\right] \sum_{q = -\infty} ^{-1} B\left(q, p\right) H_q\left(\ket{\Psi}\right).
\end{align}
For $H$ we can calculate the evolution in the same manner as for $F$. As the blue marker evolves independent of all other markers the result is however trivial:
\begin{align}
	H_p\left(M \ket{\Psi}\right) &= \sum_{q=-\infty}^\infty B\left(q, p\right) H_q\left(\ket{\Psi}\right).
\end{align}

\subsection{Proof by Induction}
\label{proofInduction}
We will now prove statement \eqref{eq:proofToShow}
\begin{align}
	\mean{F_p}_t = F_p\left(\ket{\Psi}\right)_t = H_p\left(\ket{\Psi}\right)_t = \mean{H_p}_t \text{, for } p\geq0
\end{align}
by induction. Our induction begins at 
\begin{align}
	F_p\left(\ket{\Psi}_0\right) = H_p\left(\ket{\Psi}_0\right),
\end{align}
which means that the markers have to be initialized properly for the 0-th time step. 

We will use that the blue and green markers advance identically because they follow the same equation of motion:
\begin{align}
	G\left(q, p\right) = B\left(q, p\right) \text{, for } p\geq0.
\end{align}
Our induction step is then
\begin{align}
	\label{eq:proofInductionStep}
	F_p\left(\ket{\Psi}_{t+1}\right) &= F_p\left(M \ket{\Psi}_{t}\right) \\
	&= \sum_{q=0}^\infty G\left(q, p\right) F_q\left(\ket{\Psi}_t\right) + \sum_{q = -\infty} ^{-1} B\left(q, p\right) H_q\left(\ket{\Psi}_t\right) \\
	&= \sum_{q=0}^\infty B\left(q, p\right) F_q\left(\ket{\Psi}_t\right) + \sum_{q = -\infty} ^{-1} B\left(q, p\right) H_q\left(\ket{\Psi}_t\right) \\
	&= \sum_{q=0}^\infty B\left(q, p\right) H_q\left(\ket{\Psi}_t\right) + \sum_{q = -\infty} ^{-1} B\left(q, p\right) H_q\left(\ket{\Psi}_t\right) \\
	&= \sum_{q = -\infty} ^{\infty} B\left(q, p\right) H_q\left(\ket{\Psi}_t\right) \\
	&= H_p\left(\ket{\Psi}_{t+1}\right).
\end{align}

This concludes our proof: The mean result of the simulation is the same as with no reweighting when the deterministic roulette is used. There is no bias.










\end{document}